\documentclass[reprint,amsmath,amssymb,aps,pra,floatfix]{revtex4-2}

\usepackage[T1]{fontenc}
\usepackage[utf8]{inputenc}
\usepackage{graphicx}
\usepackage[caption=false,font=footnotesize]{subfig}
\let\subfigure\subfloat
\usepackage{bm}
\usepackage{comment}
\usepackage{hyperref}
\usepackage{etoolbox}
\usepackage{microtype}

\begin{document}

\title{Retinal arterial velocity-proxy waveforms from differential spectral broadening in laser Doppler holography}

\author{Zacharie Auray}
\email{zauray@gmail.com}
\affiliation{Langevin Institute, CNRS, ESPCI Paris, PSL University, 1 rue Jussieu, 75005 Paris, France}

\author{Yann Fischer}
\affiliation{Langevin Institute, CNRS, ESPCI Paris, PSL University, 1 rue Jussieu, 75005 Paris, France}

\author{Olivier Martinache}
\affiliation{Langevin Institute, CNRS, ESPCI Paris, PSL University, 1 rue Jussieu, 75005 Paris, France}

\author{Michael Atlan}
\affiliation{Langevin Institute, CNRS, ESPCI Paris, PSL University, 1 rue Jussieu, 75005 Paris, France}

\date{\today}

\begin{abstract}
Laser Doppler holography (LDH) records pulsatile spectral broadening in retinal arteries. We compare vessel and local-background measurements to extract a signed differential-broadening waveform, $\Delta\sigma$. A minimal Gaussian convolution model motivates quadratic subtraction, and Doppler scaling relates the added width to a characteristic blood velocity. Real arteries show pulsatile waveforms, whereas identically processed imaginary arteries show only low-amplitude fluctuations around zero. We interpret the empirically scaled signal as a velocity proxy for waveform shape and within-vessel changes during a single acquisition under stable conditions.
\end{abstract}


\maketitle

\begin{figure}[!t]
\centering
\includegraphics[width=\linewidth]{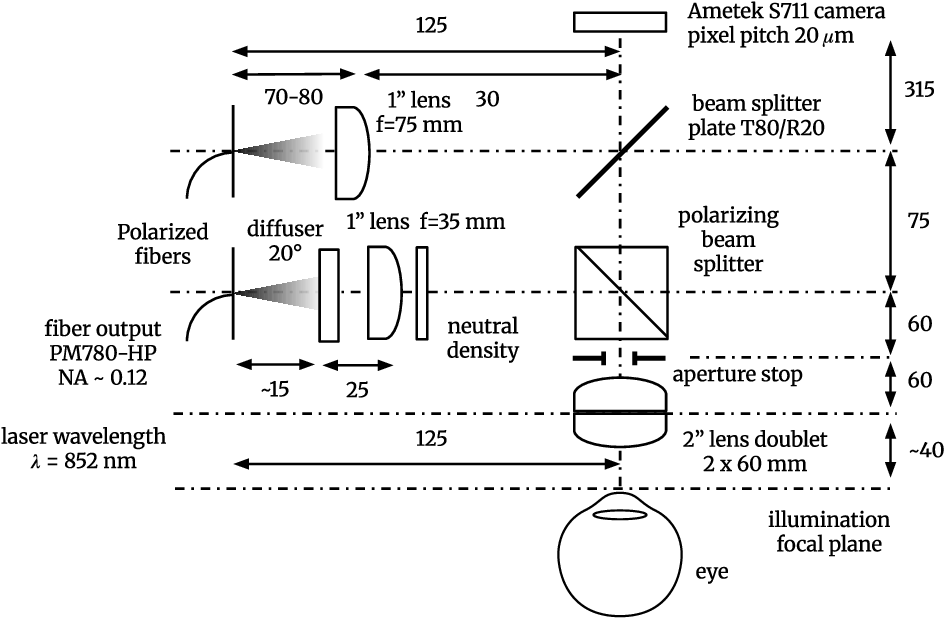}
\caption{Doppler holography setup. An inline Mach--Zehnder interferometer combines fundus return light with a reference beam.}
\label{fig:optical_setup}
\end{figure}

\par\medskip \noindent\textbf{Introduction.} Retinal arterial waveforms describe the dynamics of blood supply over the cardiac cycle. We use laser Doppler holography (LDH) to extract a signed differential-broadening waveform from paired vessel and local-background measurements. A minimal Gaussian convolution model motivates the correction; manually drawn ``imaginary arteries'' provide controls for a systematic residual after background correction.

Speckle methods infer motion from temporal light fluctuations. Intensity detection measures second-order correlations $g_2(\tau)$, while interferometric detection accesses first-order field correlations $g_1(\tau)$ and their Doppler spectra. Laser speckle contrast imaging (LSCI) measures exposure-integrated fluctuations, while multi-exposure speckle imaging (MESI) fits correlation times across several exposures~\cite{parthasarathy_robust_2008}. Both require correlation models that account for static light and instrumental averaging. Rapid estimation of multi-exposure imaging (REMI) accelerates this inversion for real-time perfusion imaging~\cite{rivera_quasi_analytic_2023}, whereas dynamic light scattering imaging (DLSI) samples temporal intensity correlations directly~\cite{postnov_dynamic_2020}.

Retinal studies resolved arterial pulsatility from intensity correlations~\cite{suzuki_measurement_1991} and compared speckle-derived velocities in human vessels with bidirectional laser Doppler velocimetry~\cite{nagahara_in_vivo_2011}. Laser speckle flowgraphy (LSFG) uses mean blur rate to characterize vessel and optic-nerve-head waveforms~\cite{enomoto_characterization_2021}. Recent LSCI work resolved arterial and venous waveforms in the mouse retina~\cite{tate_development_2026} and cortical pulse transfer~\cite{jensen_cardiac_2026}, with high-speed human retinal results reported in a preprint~\cite{auray_high_speed_2026}.

Converting these indices to velocity depends on the scattering model. In phantoms and mouse cortex, inverse speckle correlation time tracked a combination of red blood cell speed and vessel diameter~\cite{kazmi_flux_2015}. Particularly relevant here, Nadort et al. removed neighboring-tissue decorrelation by quadrature subtraction under a Gaussian correlation model~\cite{nadort_quantitative_2013}, then incorporated scattering-order and phase-function corrections and compared velocity estimates with cell tracking in chick-embryo and human microcirculation~\cite{nadort_quantitative_2016}.

First-order approaches include dynamic light scattering optical coherence tomography (DLS-OCT), which fits complex-field correlations to estimate axial and transverse velocities, with phantom validation and cortical imaging~\cite{lee_dynamic_2012}, and functional interferometric diffusing wave spectroscopy, which resolves human cerebral pulsatility as a bulk blood-flow index~\cite{zhou_functional_2021}. LDH provides wide-field retinal spectra~\cite{puyo_in_vivo_2018} and retinal and choroidal arterial and venous waveforms, including correction for a spatially averaged background~\cite{puyo_waveform_2019}.

\begin{figure}[!t]
\centering
\subfigure[]{%
  \includegraphics[width=.48\linewidth]{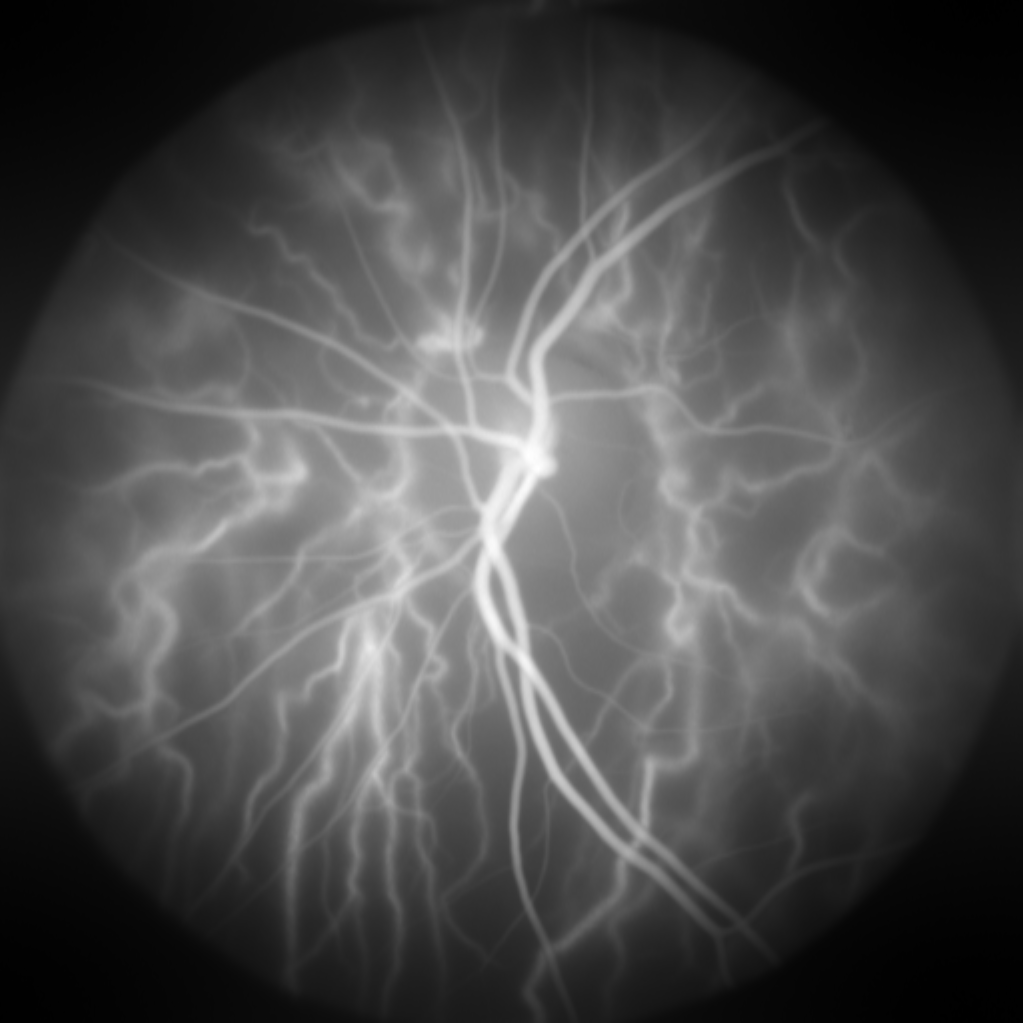}%
  \label{fig:moments_m0}%
}%
\hfill
\subfigure[]{%
  \includegraphics[width=.48\linewidth]{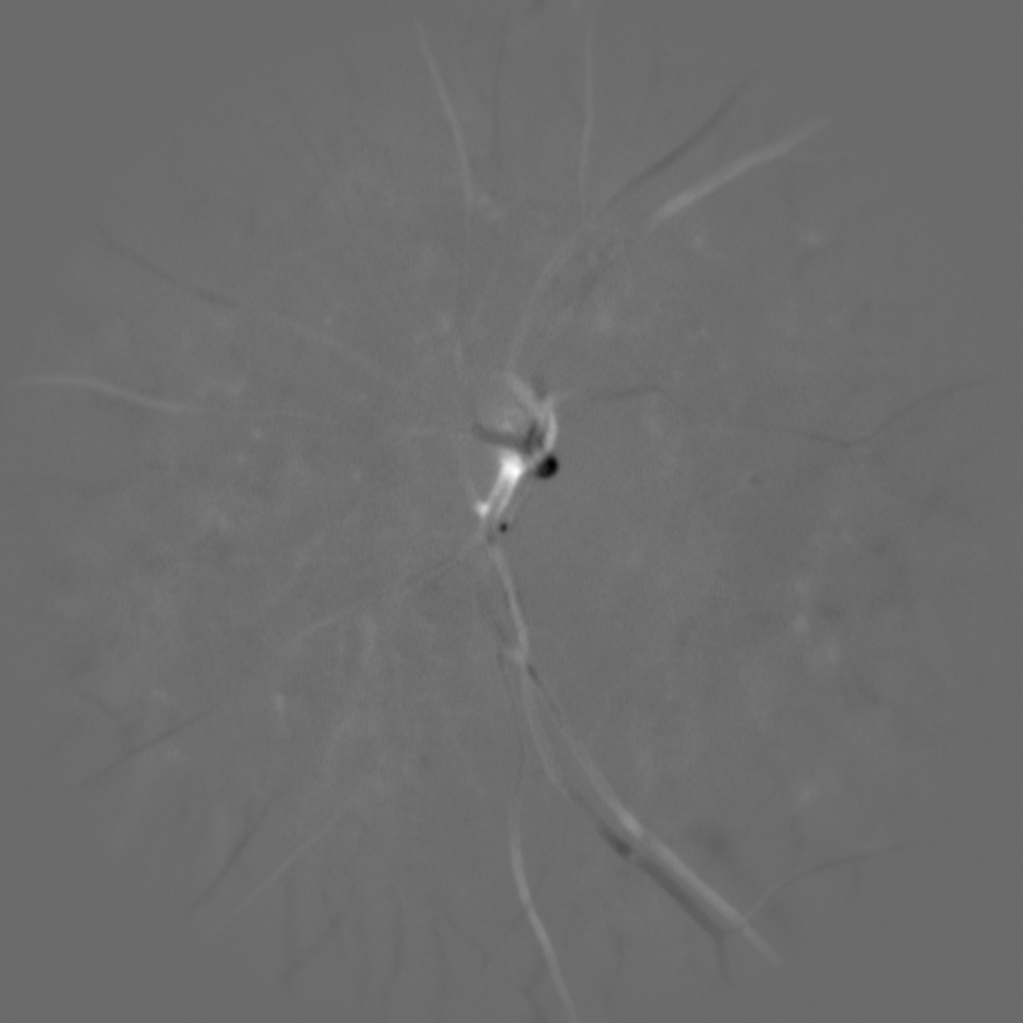}%
  \label{fig:moments_m1}%
}%
\par\vspace{2pt}
\subfigure[]{%
  \includegraphics[width=.48\linewidth]{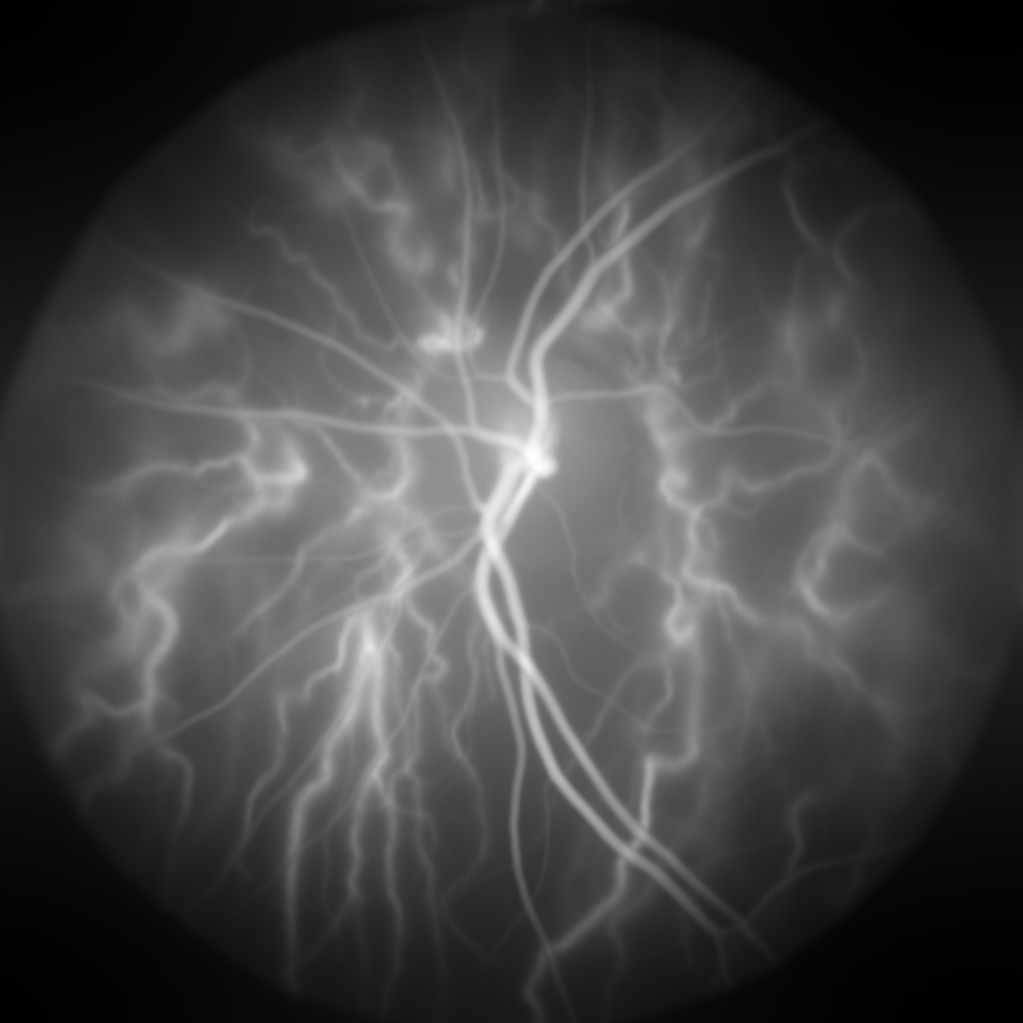}%
  \label{fig:moments_m2}%
}%
\hfill
\subfigure[]{%
  \includegraphics[width=.48\linewidth]{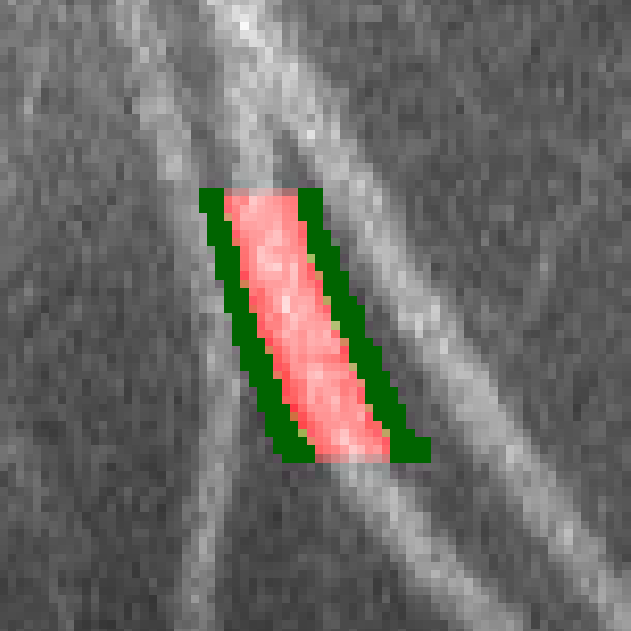}%
  \label{fig:moments_broadening}%
}%
\par\vspace{2pt}
\subfigure[]{%
  \includegraphics[width=.48\linewidth]{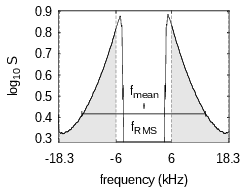}%
  \label{fig:moments_spectra}%
}%
\hfill
\subfigure[]{%
  \includegraphics[width=.48\linewidth]{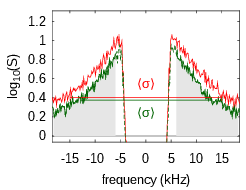}%
  \label{fig:moments_regional_widths}%
}%
\caption{Doppler moment maps and spectral broadening. \protect\subref{fig:moments_m0} Time-averaged $M_0$ map. \protect\subref{fig:moments_m1} Time-averaged $M_1$ map. \protect\subref{fig:moments_m2} Time-averaged $M_2$ map. \protect\subref{fig:moments_broadening} Local vessel and neighborhood selection. \protect\subref{fig:moments_spectra} Field-averaged Doppler spectrum, with mean frequency $f_{\mathrm{mean}}=\langle M_1\rangle/\langle M_0\rangle=0.17\,\mathrm{kHz}$ and root-mean-square frequency $f_{\mathrm{RMS}}=\sqrt{\langle M_2\rangle/\langle M_0\rangle}=13.41\,\mathrm{kHz}$, where $\langle\cdot\rangle$ denotes spatial averaging over the field of view. \protect\subref{fig:moments_regional_widths} Vessel and neighborhood spectra from \protect\subref{fig:moments_broadening}, with mean broadenings $\langle\sigma\rangle=15.00\,\mathrm{kHz}$ (red) and $12.90\,\mathrm{kHz}$ (dark green), respectively. Panels \protect\subref{fig:moments_m0}--\protect\subref{fig:moments_m2} are time-averaged over the $4.4\,\mathrm{s}$ recording. The image in \protect\subref{fig:moments_broadening} and both spectra in \protect\subref{fig:moments_spectra} and \protect\subref{fig:moments_regional_widths} are calculated from the same batch of 256 frames. The moment maps use linear grayscale displays in arbitrary units.}
\label{fig:moments}
\end{figure}

\par\medskip \noindent\textbf{Image acquisition and processing.} We used the previously described LDH system, diffuse illumination, and real-time rendering workflow~\cite{fischer_retinal_2024,bratasz_diffuse_2024,puyo_in_vivo_2018}. The Mach--Zehnder inline interferometer was illuminated by an $852\,\mathrm{nm}$ diode laser (Thorlabs FPV852P). Polarization-maintaining fibers (Thorlabs PM780-HP) split the beam into linearly polarized reference and illumination arms with a 10/90 ratio. An engineered diffuser (Thorlabs ED1-C20-MD, 1'' diameter, $20^\circ$ circular top-hat) shaped the illumination beam. An eyepiece comprising two biconvex lenses, with an effective focal length of $\sim33\,\mathrm{mm}$, focused the beam onto the eye. To maximize retinal illumination, the focus was placed near the eye's optical center, with the eye-to-doublet distance close to the doublet focal length. Cross-polarized fundus return light interfered with the reference beam. An Ametek Phantom S711 camera recorded the interferograms at $37.037\,\mathrm{kHz}$ over $512\times320$ pixels with a $20\,\mu\mathrm{m}$ pixel pitch~\cite{auray_ultrahigh_2025}. Frames were transferred through Euresys Coaxlink QSFP+ frame grabbers and saved using Holovibes v14.11.12. Exposure time was $24\,\mu\mathrm{s}$ for a $27\,\mu\mathrm{s}$ frame period. Illumination power was $5\,\mathrm{mW}$, measured at the focus of the eyepiece/relay lens.

Measurements were obtained from one eye of one volunteer, without pharmacological pupil dilation. A standard chinrest provided head support. The field was centered on the optic nerve head (ONH) to record proximal branches of the central retinal artery simultaneously. The volunteer gave informed consent, and the measurements followed the principles of the Declaration of Helsinki. Each recording comprised 163840 raw holograms spanning approximately $4.4\,\mathrm{s}$.

Offline reconstruction used a Fresnel transformation with a propagation distance of $480\,\mathrm{mm}$. The $512\times320$ interferograms were resized to obtain a $512\times512$ grid with an approximately $20\,\mu\mathrm{m}$ reconstructed pixel size on the retina. In the inline configuration (Fig.~\ref{fig:optical_setup}), the conjugate image and self-beating terms overlap the retinal reconstruction. These blurred contributions form a uniform baseline across the field.

Singular value decomposition (SVD) filtering suppressed spatially coherent clutter, including eye motion and corneal reflections~\cite{puyo_spatio-temporal_2020}. We filtered the complex Fresnel-propagated frames in non-overlapping blocks of 256 frames, rejecting the 64 components associated with the largest singular values. A short-time Fourier transform (STFT) within 256-frame windows yielded Doppler spectra with frequency expressed in Hz. Components below $6\,\mathrm{kHz}$ in absolute frequency were excluded to reduce bulk eye motion and slow background fluctuations.

The spectral moments of the Doppler power spectrum $S(x,y,t,f)$ were calculated pixel by pixel for orders $n=0,1,2$:

\begin{equation}
M_n (x,y,t)= \int _\Omega S(x,y,t,f) f^n {\, \rm d} f
\label{eq:moments}
\end{equation}

Here, $\Omega=[-f_2,-f_1]\cup[f_1,f_2]$, with $f_1=6\,\mathrm{kHz}$ and $f_2=18.3\,\mathrm{kHz}$, slightly below the Nyquist frequency. The moment maps were registered by translation after moment calculation. Two-dimensional cross-correlation of $M_0$ images against an $M_0$ reference estimated the translations, which were applied identically to all three maps. Quantitative masks and profiles were evaluated on the original $512\times512$ reconstruction grid. The $M_0$ and $M_2$ maps show retinal and choroidal vessels. The $M_1$ map highlights spectral asymmetry associated with velocity components normal to the retinal plane, notably in the central retinal artery and vein (Fig.~\ref{fig:moments}).

Using $M_1$, $M_2$, and the field-averaged zeroth moment $\langle M_0 \rangle_{x,y}$, we define a signed first-moment contrast $\mu(x,y,t)$ and a second-moment broadening contrast $\sigma(x,y,t)$.

\begin{equation}
\mu (x, y, t)= \dfrac{M_1(x, y, t)}{\langle M_0 \rangle_{x, y}(t)},\quad
\sigma (x, y, t)= \sqrt{\dfrac{M_2(x, y, t)}{\langle M_0 \rangle_{x, y}(t)}}
\label{eq:fRMS}
\end{equation}

Frequency $f$ is expressed in Hz throughout the spectral calculations; $\mu$, $\sigma$, and $\Delta\sigma$ therefore have units of Hz.

Spatial averaging of $M_0$ over the full field of view improves the signal-to-noise ratio of the denominator and stabilizes $\mu$ and $\sigma$. With this shared normalization, the two quantities are spectral contrasts rather than the centroid and standard deviation of a locally normalized spectrum. The contrast $\sigma$ depends on both local spectral power and spectral shape over the selected band, so its interpretation as a spectral width is approximate.

Rendering with \href{https://github.com/DigitalHolography/HoloDopplerMatlab/tree/2.9.12}{\textit{HoloDoppler v2.9.12}} (revision \href{https://github.com/DigitalHolography/HoloDopplerMatlab/commit/5c12622a92ed19a35f93afa71368c74837a3fb40}{\texttt{5c12622}}) generated 639-frame videos of the three moments at a time step of approximately $6.9\,\mathrm{ms}$. The plotted time axis starts at the first rendered frame. Subsequent analysis used \href{https://github.com/DigitalHolography/EyeFlowMatlab}{EyeFlow Matlab v2.10.1}, with the reference source preserved at revision \href{https://github.com/DigitalHolography/EyeFlowMatlab/commit/fff92d96abb0084de2c63c797c8ffbad78581625}{\texttt{fff92d9}}.

\begin{figure}[!t]
\centering
\makebox[.48\linewidth][c]{\fontsize{7}{8.4}\selectfont\bfseries retinal arteries}%
\hfill
\makebox[.48\linewidth][c]{\fontsize{7}{8.4}\selectfont\bfseries imaginary retinal arteries}%
\par\vspace{2pt}
\subfigure[]{%
  \includegraphics[width=.48\linewidth]{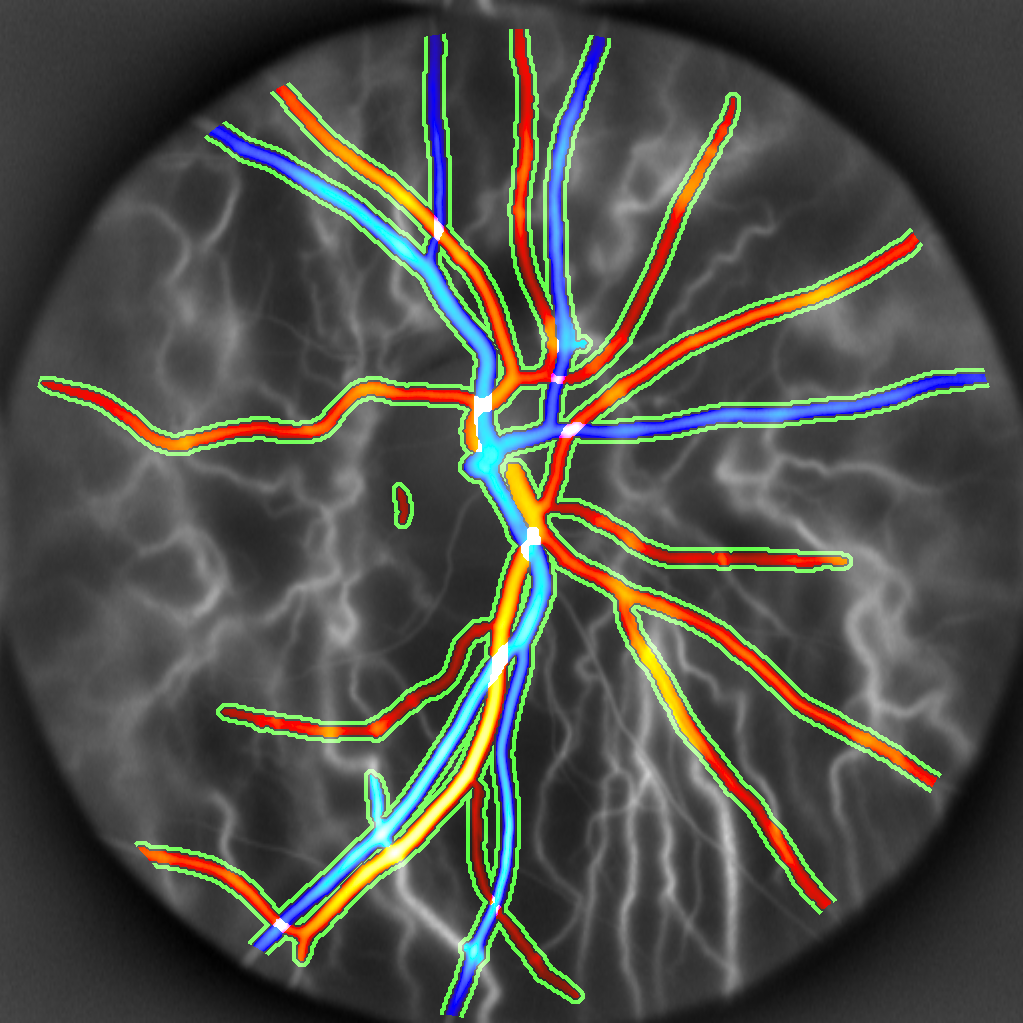}%
  \label{fig:controls_artery_mask}%
}%
\hfill
\subfigure[]{%
  \includegraphics[width=.48\linewidth]{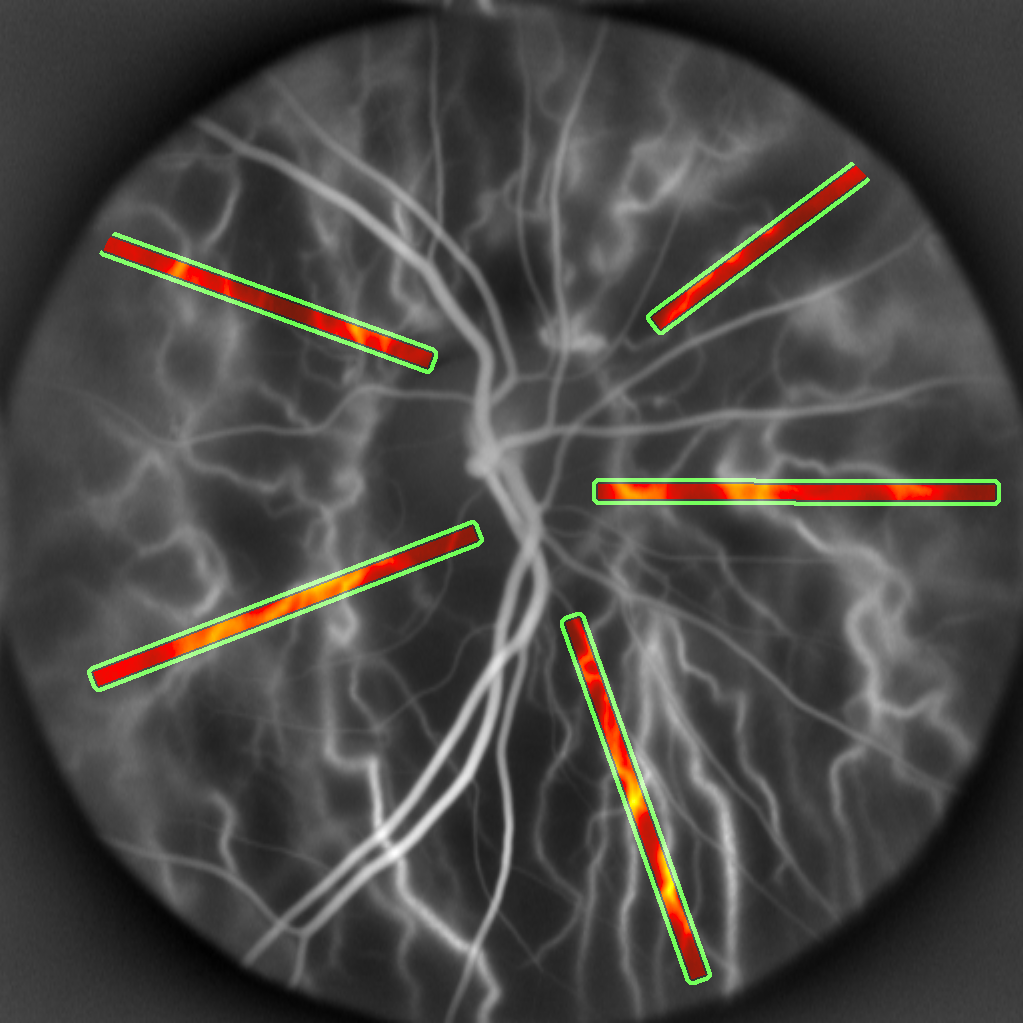}%
  \label{fig:controls_reference_mask}%
}%
\par\vspace{2pt}
\subfigure[]{%
  \includegraphics[width=.48\linewidth]{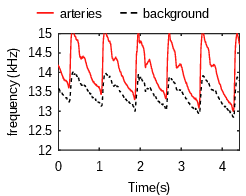}%
  \label{fig:controls_artery_spectral}%
}%
\hfill
\subfigure[]{%
  \includegraphics[width=.48\linewidth]{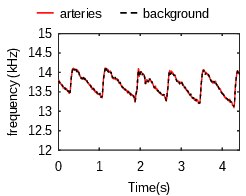}%
  \label{fig:controls_reference_spectral}%
}%
\par\vspace{2pt}
\subfigure[]{%
  \includegraphics[width=.48\linewidth]{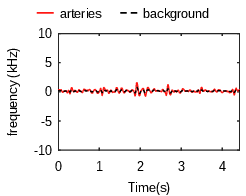}%
  \label{fig:controls_artery_axial}%
}%
\hfill
\subfigure[]{%
  \includegraphics[width=.48\linewidth]{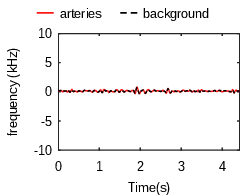}%
  \label{fig:controls_reference_axial}%
}%
\caption{Real and imaginary arteries processed identically. \protect\subref{fig:controls_artery_mask},\protect\subref{fig:controls_reference_mask} Real- and imaginary-artery masks. \protect\subref{fig:controls_artery_spectral},\protect\subref{fig:controls_reference_spectral} Spatial means of $\sigma$ and its local background. \protect\subref{fig:controls_artery_axial},\protect\subref{fig:controls_reference_axial} Spatial means of $\mu$ and its local background. Red and dashed black traces denote selected masks and backgrounds, respectively; frequencies are in kHz. Both columns use the same annular section.}
\label{fig:realVsImagArteries}
\end{figure}

\begin{figure}[!t]
\centering
\makebox[.48\linewidth][c]{\fontsize{7}{8.4}\selectfont\bfseries retinal arteries}%
\hfill
\makebox[.48\linewidth][c]{\fontsize{7}{8.4}\selectfont\bfseries imaginary retinal arteries}%
\par\vspace{2pt}
\subfigure[]{%
  \includegraphics[width=.48\linewidth]{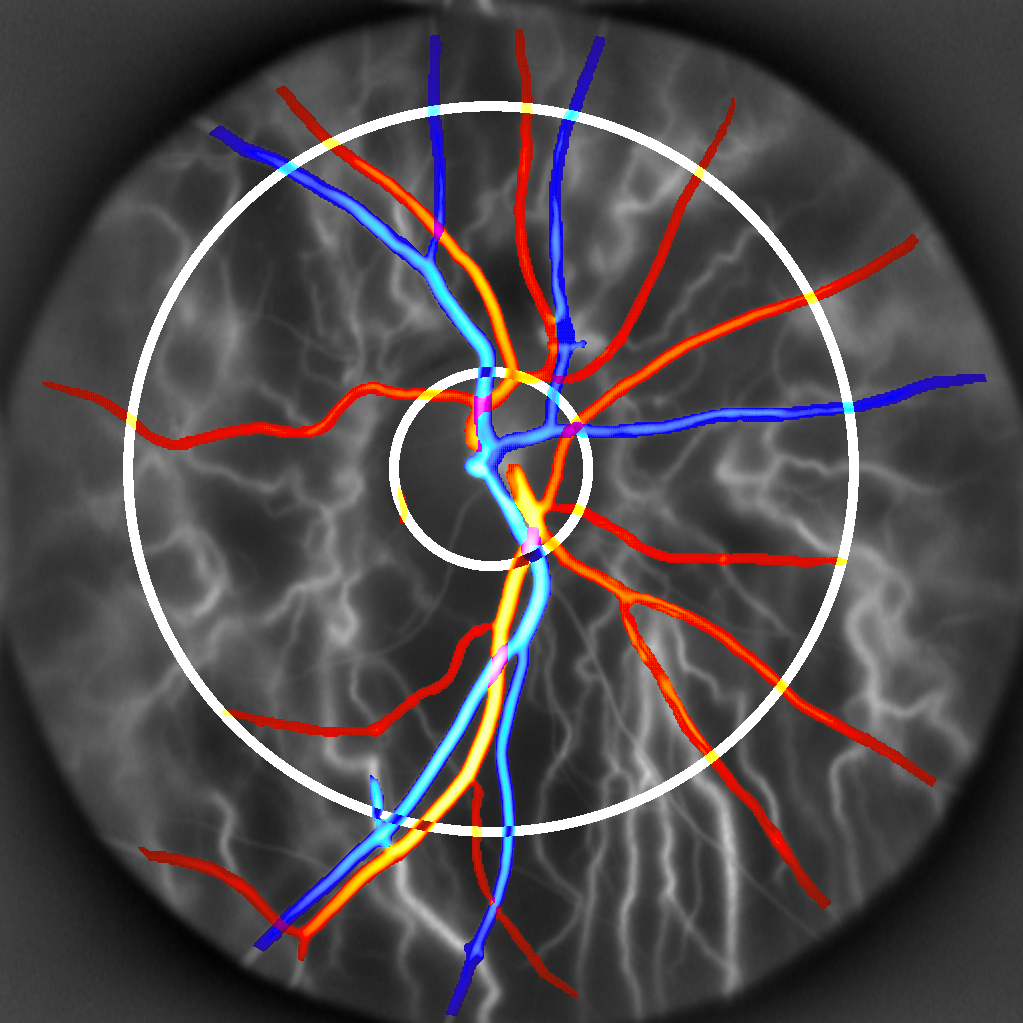}%
  \label{fig:sections_artery}%
}%
\hfill
\subfigure[]{%
  \includegraphics[width=.48\linewidth]{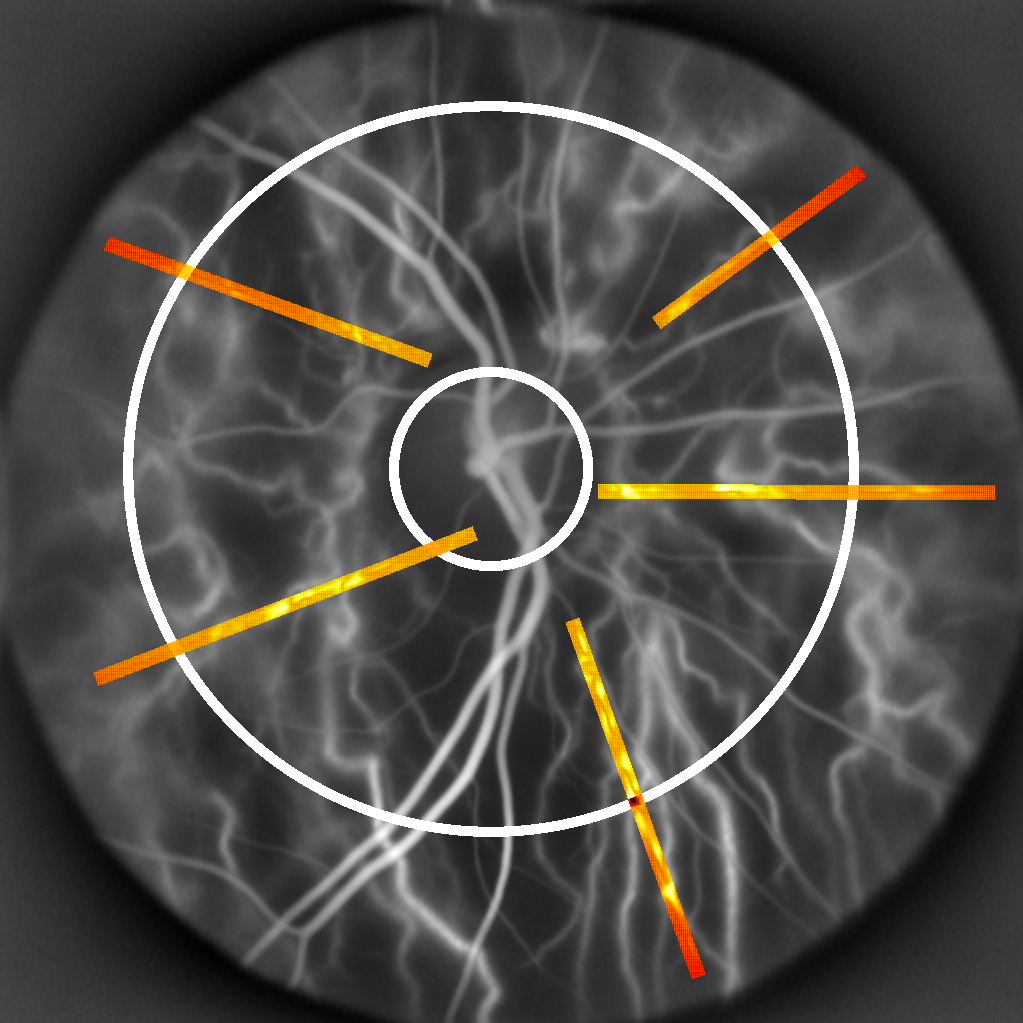}%
  \label{fig:sections_reference}%
}%
\par\vspace{2pt}
\subfigure[]{%
  \includegraphics[width=.48\linewidth]{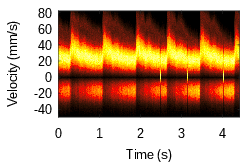}%
  \label{fig:sections_artery_distribution}%
}%
\hfill
\subfigure[]{%
  \includegraphics[width=.48\linewidth]{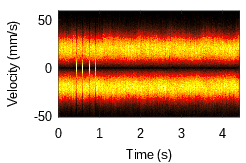}%
  \label{fig:sections_reference_distribution}%
}%
\par\vspace{2pt}
\subfigure[]{%
  \includegraphics[width=.48\linewidth]{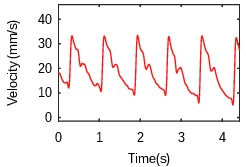}%
  \label{fig:sections_artery_waveform}%
}%
\hfill
\subfigure[]{%
  \includegraphics[width=.48\linewidth]{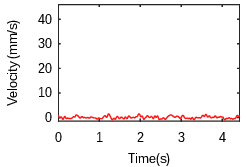}%
  \label{fig:sections_reference_waveform}%
}%
\caption{Real-artery (left) and imaginary-artery (right) velocity proxies. \protect\subref{fig:sections_artery},\protect\subref{fig:sections_reference} Sections bounded by two circles. \protect\subref{fig:sections_artery_distribution},\protect\subref{fig:sections_reference_distribution} Proxy distributions. \protect\subref{fig:sections_artery_waveform},\protect\subref{fig:sections_reference_waveform} Section-averaged waveforms. The imaginary-artery waveform shows only small fluctuations around zero compared with the pulsatile real-artery signal, providing a negative control for residual background signals at the selected locations. The mm/s axes use the empirical scaling in Eq.~(\ref{eq:local_velocity_estimate}).}
\label{fig:realVsImagArteries2}
\end{figure}

Masks for real arteries and five imaginary arteries were drawn manually; adjacent neighborhoods provided paired background measurements (Fig.~\ref{fig:realVsImagArteries}). The neighborhoods directly bordered the masks, with no gap or special handling at crossings. At each frame, background broadening was reconstructed with MATLAB \texttt{inpaintCoherent}; background first-moment contrasts were estimated by masked neighborhood averaging. Waveforms were formed from arithmetic means over masked pixels in the ONH-centered annular section bounded by the two circles in Fig.~\ref{fig:realVsImagArteries2}\subref{fig:sections_artery},\subref{fig:sections_reference}, with inner and outer radii of 0.10 and 0.35 times the image width, respectively. For the signed differential broadening defined in Eq.~(\ref{eq:signed_broadening}), the signed square root was applied pixel by pixel before spatial averaging.

The reference workflow detects outlier frames with a 10-frame moving-median window and replaces them by linear interpolation before filtering. The derived waveforms were filtered with a fourth-order Butterworth low-pass filter at $15\,\mathrm{Hz}$, applied forward and backward.

\par\medskip \noindent\textbf{Differential broadening as a velocity proxy.} At $852\,\mathrm{nm}$, light returning from the deeper fundus includes contributions from the retinal pigment epithelium (RPE), choroid, and sclera~\cite{delori_spectral_1989,vandekraats_pathways_1996,vogel_sclera_1991}. This light may already be Doppler broadened before crossing a superficial retinal vessel~\cite{puyo_choroidal_2019}. We treat it as a local secondary field that gains additional broadening through interactions with moving red blood cells.

For approximately in-plane vessels with nearly symmetric spectra, we model both the background spectrum and the additional vessel-induced shifts as Gaussian. Let $G(f)$ be the normalized, zero-mean distribution of the added shifts, with standard deviation $s$ in Hz. The normalized vessel and background spectra, $G_{\mathrm{RV}}(f)$ and $G_{\mathrm{BKG}}(f)$, have standard deviations $s_{\mathrm{RV}}$ and $s_{\mathrm{BKG}}$. If neighboring tissue represents the background beneath the vessel, and the added shifts are independent of the background broadening, then before frequency selection
\begin{equation}
G_{\mathrm{RV}}(f)
=(G_{\mathrm{BKG}}*G)(f).
\label{eq:gaussian_convolution}
\end{equation}
Gaussian variances add exactly, giving the spectral counterpart of Gaussian decorrelation corrections~\cite{nadort_quantitative_2013}:
\begin{equation}
s=\sqrt{s_{\mathrm{RV}}^2-s_{\mathrm{BKG}}^2}.
\label{eq:gaussian_width_difference}
\end{equation}
This model motivates the quadratic correction below; Gaussian fits were not performed. The Gaussian assumption applies to Doppler frequency shifts and does not require Gaussian-distributed red blood cell speeds.

For one scattering event, $\Delta f=(\Delta\mathbf{k}\cdot\mathbf{v})/(2\pi)$, where $\Delta\mathbf{k}=\mathbf{k}_s-\mathbf{k}_i$. For zero-mean added shifts, we assume a fixed distribution of scattering directions and a velocity distribution whose shape remains constant as its amplitude varies. Then
\begin{equation}
s^2=\frac{1}{(2\pi)^2}\left\langle(\Delta\mathbf{k}\cdot\mathbf{v})^2\right\rangle_{\theta,\phi}
\propto\frac{v^2}{\lambda^2}.
\label{eq:doppler_scaling}
\end{equation}
Here, $\langle\cdot\rangle_{\theta,\phi}$ denotes an angular average over the effective scattering geometry with normalized optical weights, $v$ is the characteristic velocity amplitude, and $\lambda$ is the vacuum wavelength. The wave-vector transfer $\Delta\mathbf{k}$ includes both incident and scattered directions. At fixed wavelength and angular weighting, $s$ therefore scales with $v$.

\begin{figure}[!t]
\centering
\makebox[.48\linewidth][c]{\fontsize{7}{8.4}\selectfont\bfseries retinal arteries}%
\hfill
\makebox[.48\linewidth][c]{\fontsize{7}{8.4}\selectfont\bfseries imaginary retinal arteries}%
\par\vspace{2pt}
\subfigure[]{%
  \shortstack[c]{%
    \includegraphics[width=.48\linewidth]{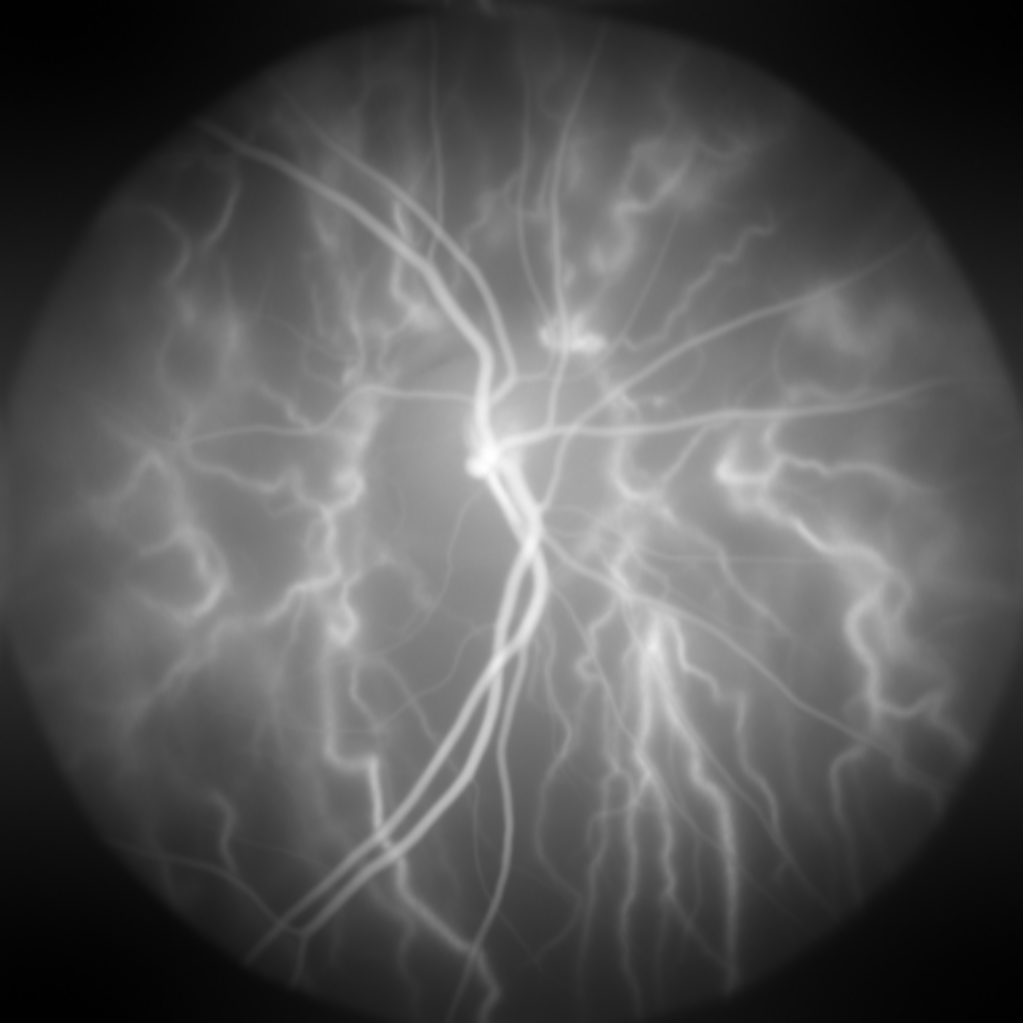}\\[-2pt]
    {\fontsize{6}{7}\selectfont $\langle\sigma\rangle_t$ (kHz)}\\[-2pt]
    \includegraphics[width=.3936\linewidth,trim=0 0 0 23.04bp,clip]{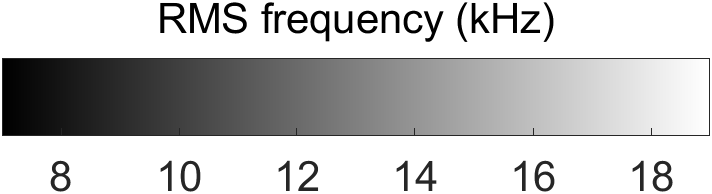}
  }%
  \label{fig:broadening_vessel}%
}%
\hfill
\subfigure[]{%
  \shortstack[c]{%
    \includegraphics[width=.48\linewidth]{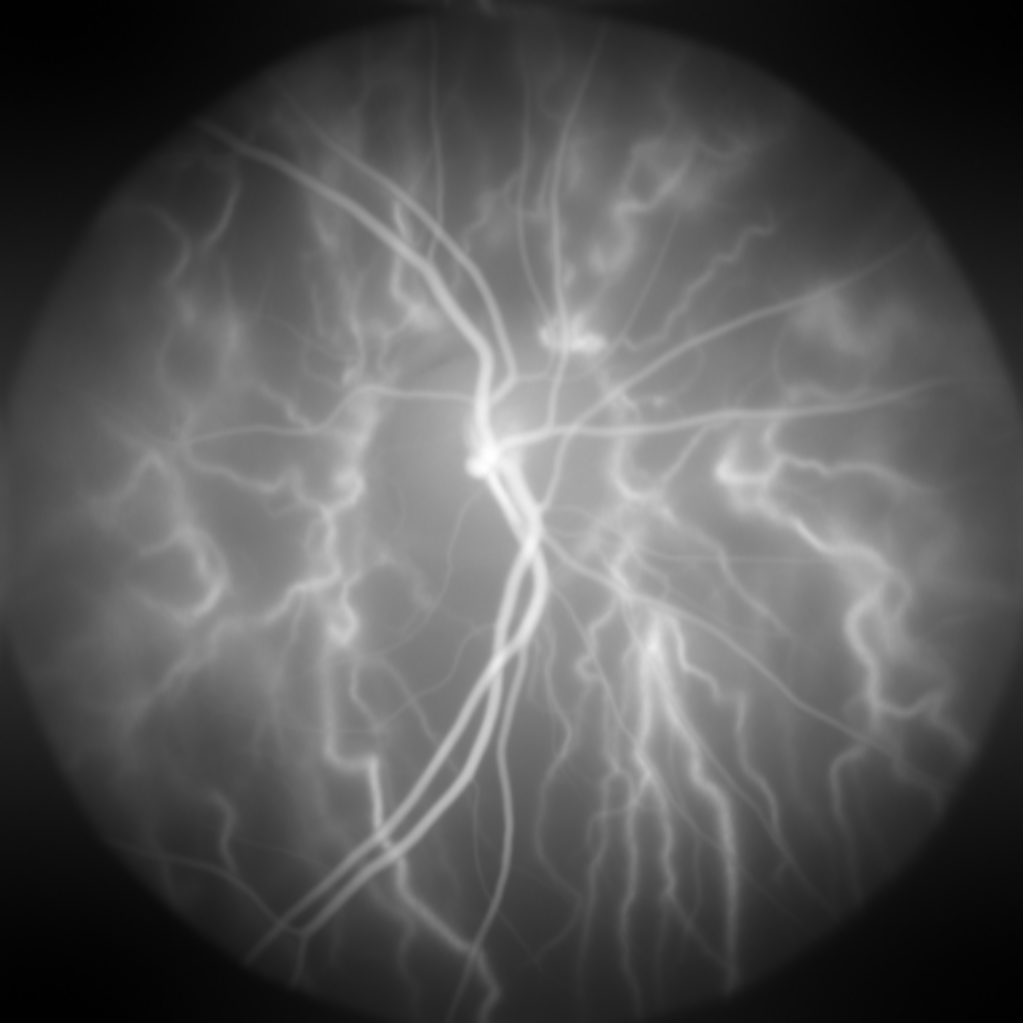}\\[-2pt]
    {\fontsize{6}{7}\selectfont $\langle\sigma\rangle_t$ (kHz)}\\[-2pt]
    \includegraphics[width=.3936\linewidth,trim=0 0 0 23.04bp,clip]{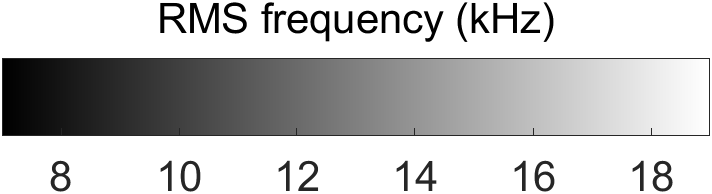}
  }%
  \label{fig:broadening_reference}%
}%
\par\vspace{2pt}
\subfigure[]{%
  \shortstack[c]{%
    \includegraphics[width=.48\linewidth]{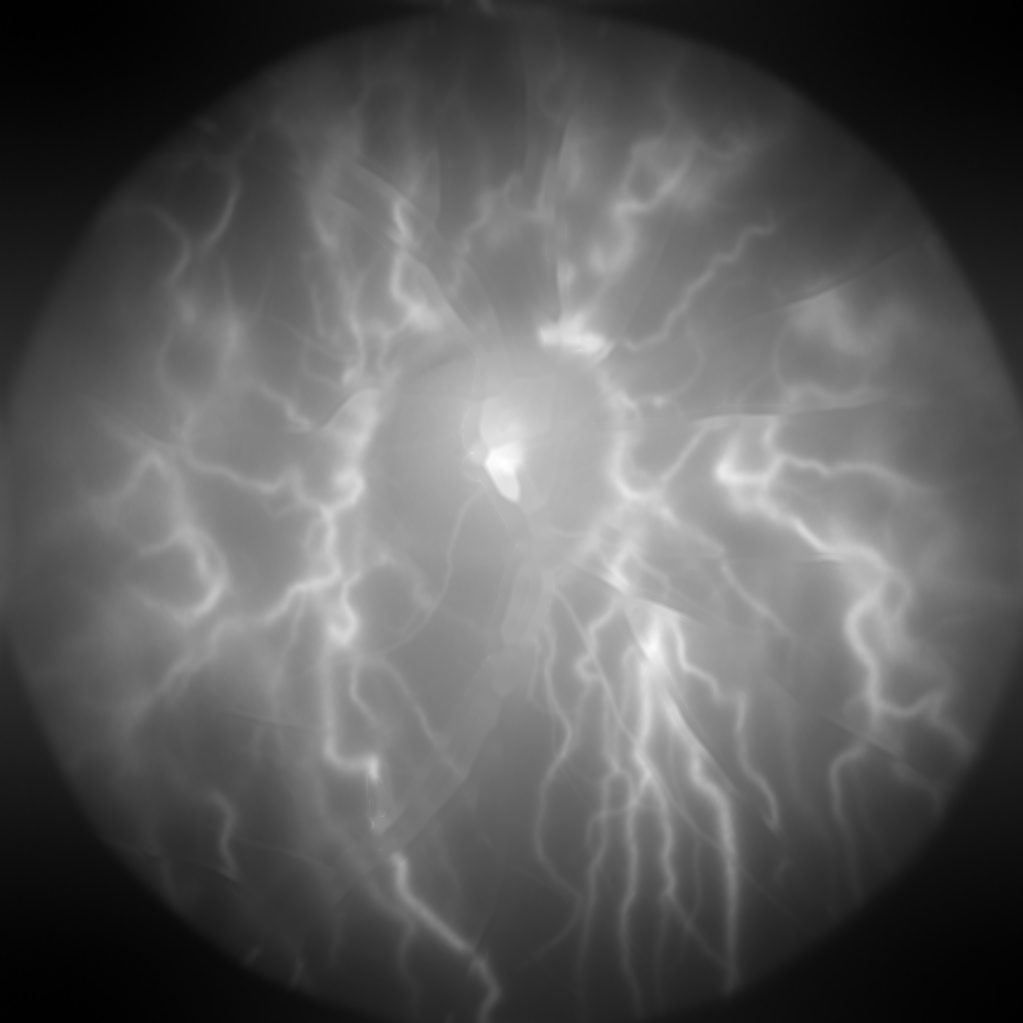}\\[-2pt]
    {\fontsize{6}{7}\selectfont $\langle\sigma_{\mathrm{BKG}}\rangle_t$ (kHz)}\\[-2pt]
    \includegraphics[width=.3936\linewidth,trim=0 0 0 23.04bp,clip]{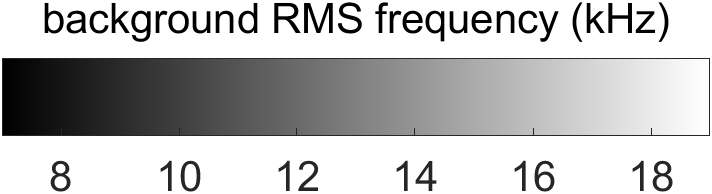}
  }%
  \label{fig:broadening_vessel_background}%
}%
\hfill
\subfigure[]{%
  \shortstack[c]{%
    \includegraphics[width=.48\linewidth]{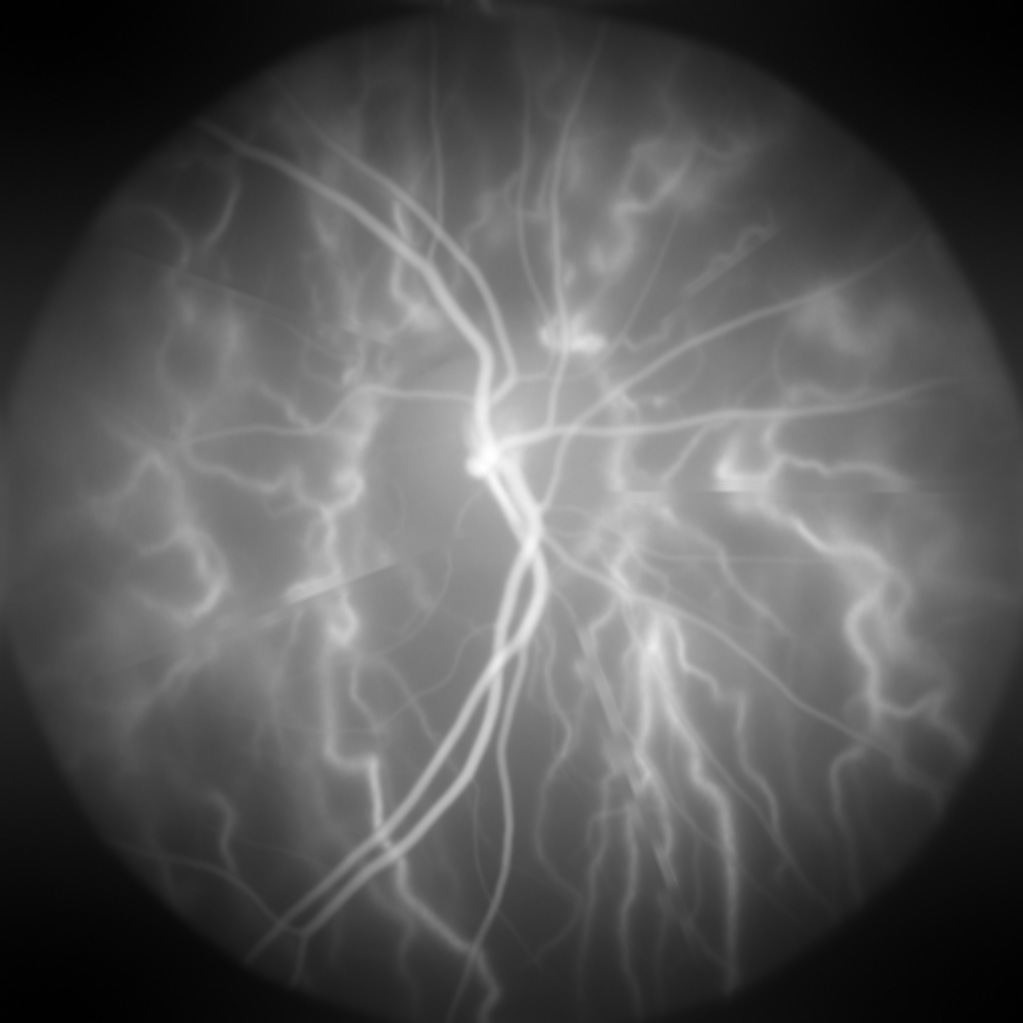}\\[-2pt]
    {\fontsize{6}{7}\selectfont $\langle\sigma_{\mathrm{BKG}}\rangle_t$ (kHz)}\\[-2pt]
    \includegraphics[width=.3936\linewidth,trim=0 0 0 23.04bp,clip]{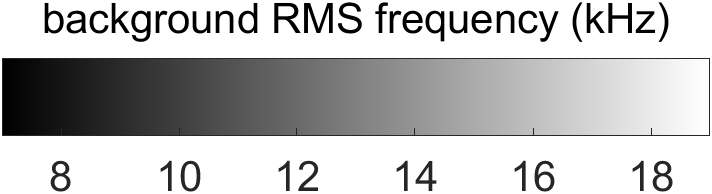}
  }%
  \label{fig:broadening_reference_background}%
}%
\caption{Time-averaged broadening and reconstructed backgrounds for real and imaginary arteries. \protect\subref{fig:broadening_vessel} Broadening $\langle\sigma\rangle_t$ for the real-artery analysis. \protect\subref{fig:broadening_reference} Broadening for the imaginary-artery analysis. \protect\subref{fig:broadening_vessel_background} Reconstructed real-artery background $\langle\sigma_{\mathrm{BKG}}\rangle_t$. \protect\subref{fig:broadening_reference_background} Reconstructed imaginary-artery background. Map intensities are linearly rescaled for display; the accompanying frequency scales are in kHz. Differential broadening is calculated frame by frame before temporal averaging.}
\label{fig:broadening_comparison}
\end{figure}

Equation~(\ref{eq:gaussian_width_difference}) applies to normalized Gaussian spectra. To correct the measured broadening contrasts $\sigma$ from Eq.~(\ref{eq:fRMS}), we use the same quadratic form and define signed differential broadening as
\begin{equation}
\Delta \sigma(x,y,t)
=
\mathrm{sign}
\left[
\sigma_{\mathrm{RV}}^2-
\sigma_{\mathrm{BKG}}^2
\right]
\sqrt{
\left|
\sigma_{\mathrm{RV}}^2-
\sigma_{\mathrm{BKG}}^2
\right|
}.
\label{eq:signed_broadening}
\end{equation}
The sign preserves negative differences during averaging. A negative value means that the local contrast is below its background estimate; it does not indicate reversed blood flow. Equation~(\ref{eq:gaussian_width_difference}) describes the quadratic subtraction of variances for normalized Gaussian spectra. Our experimental estimator instead uses band-limited second moments normalized by the field-averaged spectral power. Consequently, $\Delta\sigma$ depends on both local spectral broadening and spectral power, and cannot be identified directly with the standard deviation of the additional Doppler shifts. We interpret it as a signed differential spectral contrast, approximately proportional to blood velocity under scattering conditions requiring further validation.

Using the working approximation $s\propto\Delta\sigma$ over the selected band with Eq.~(\ref{eq:doppler_scaling}), we write
\begin{equation}
v_{\mathrm{proxy}}(t)=C\lambda\Delta\sigma(t),
\label{eq:local_velocity_estimate}
\end{equation}
Here, $C=10$ is the empirical scaling coefficient used in the MATLAB scripts for the reported results. It was chosen to place retinal arterial velocity estimates within the range typically reported in healthy controls~\cite{riva_blood_1985,zhong_noninvasive_2011}. At $852\,\mathrm{nm}$, this gives $8.5\,\mathrm{mm\,s^{-1}}$ per kHz of differential broadening. This empirical scaling does not independently validate absolute velocity.

\par\medskip \noindent\textbf{Real and imaginary arteries.} Real arteries show pulsatile waveforms; identically processed imaginary arteries show low-amplitude fluctuations around zero (Figs.~\ref{fig:realVsImagArteries}--\ref{fig:realVsImagArteries2}). These control signals remain small after local background correction, providing a negative control for residual background signals at the selected locations.

\par\medskip \noindent\textbf{Discussion.} We use the velocity proxy to assess retinal blood-flow waveform shape through metrics invariant to a constant positive velocity gain, including normalized timing and dimensionless pulsatility ratios~\cite{buisson_transportable_2026}. For valid arterial waveforms, a gain constant over the analyzed beat rescales amplitude without changing these metrics. For this field-normalized proxy, the conversion factor depends on the weighted incident and collected directions, the global and local illumination power before normalization, blood optical properties including hematocrit, vessel geometry, and interactions with moving red blood cells~\cite{bonner_model_1981,jentink_monte_1990}. Rather than a single empirical constant, $C$ could be modeled as a function of local retinal optical and geometrical properties, together with the illumination and collection geometry, to improve velocity estimation.

Gain invariance does not compensate for waveform distortion. Cardiac changes in caliber, scattering, or alignment can modify waveform shape. The broadening contrast also depends on the selected spectral band, noise treatment, field normalization, SVD filtering, segmentation, and background interpolation. Comparing fitted Gaussian widths with $\Delta\sigma$ over the processing band would test the approximation. Controlled-flow experiments, repeated retinal acquisitions, and comparison with established retinal velocimetry~\cite{riva_blood_1985} are needed to establish this estimator's operating range, repeatability, and independent velocity calibration, including validation of a locally varying $C$. Volumetric-flow estimation requires validated cross-sectional mean velocity and lumen area.

\par\medskip \noindent\textbf{Conclusion.} LDH provides wide-field, time-resolved retinal vascular waveforms. Previous LDH work using a related differential-broadening estimator demonstrated repeatability of retinal arterial hemodynamic metrics~\cite{martinache_repeatability_2026}, supporting further evaluation for functional vascular assessment. Independent calibration of absolute velocity remains necessary. Imaginary arteries provide a low-amplitude negative control. This work lays the foundation for using dynamic light scattering models based on forward scattering to assess blood velocity in retinal vessels.

\par\medskip \noindent\textbf{Funding.} ANR LIDARO, ANR-22-CE19-0033-01.

\par\medskip \noindent\textbf{Disclosures.} The authors declare no conflicts of interest.

\par\medskip \noindent\textbf{Data availability.} Raw interferograms and visualizations are available on \href{https://doi.org/10.5281/zenodo.24279836}{Zenodo}.

\makeatletter
\patchcmd{\bibsection}{\addvspace{19\p@}}{\addvspace{6\p@}}{}{}
\makeatother
\bibliographystyle{apsrev4-2}
\bibliography{velocity}

\end{document}